\documentclass[conference, letter]{IEEEtran}
\IEEEoverridecommandlockouts

\ifCLASSINFOpdf
  \usepackage[pdftex]{graphicx}
  \DeclareGraphicsExtensions{.pdf,.jpeg,.png}
\else
\fi

\usepackage{cite}
\usepackage{amsmath,amssymb,amsfonts}

\usepackage{algorithm} 
\usepackage{algpseudocode} 
\usepackage{graphicx}
\usepackage{textcomp}
\usepackage[left=1.29cm, right=1.29cm, top=1.9cm, bottom=4.29cm]{geometry}
\usepackage{enumitem}
\usepackage{anyfontsize}
\usepackage{fancyhdr}
\makeatletter
\renewcommand{\fnum@figure}{Figure \thefigure}
\makeatother

\title{ {Fleet Size and Spill for UAM Operation under Uncertain Demand} \\
\vspace{0.5cm}
\thanks{*This research was funded by Supernal, grant number 052838-002.}
}

\author{\IEEEauthorblockN{Shangqing Cao, Xuan Jiang, Emin Burak Onat, \\Mark Hansen, Raja Sengupta}
\IEEEauthorblockA{University of California, Berkeley \\
Berkeley, CA, USA}
\and
\IEEEauthorblockN{Bo Zou}
\IEEEauthorblockA{University of Illinois, Chicago \\
Chicago, IL, USA}  
\and

\IEEEauthorblockN{Anjan Chakrabarty}
\IEEEauthorblockA{Supernal \\
Fremont, CA, USA}  
}

\renewcommand\IEEEkeywordsname{Keywords}
\IEEEaftertitletext{\vspace{-1\baselineskip}}

\begin{document}

\maketitle

\thispagestyle{fancy}

\noindent \begin{abstract}
Variation and imbalance in demand poses significant challenges to Urban Air Mobility (UAM) operations, affecting strategic decisions such as fleet sizing. To study the implications of demand variation on UAM fleet operations, we propose a stochastic passenger arrival time generation model that uses real-world data to infer demand distributions, and two integer programs that compute the zero-spill fleet size and the spill-minimizing flight schedules and charging policies, respectively. Our numerical experiment on a two-vertiport network shows that spill in relatively inelastic to fleet size and that the driving factor behind spill is the imbalance in demand.
\end{abstract}

\vspace{0.3cm}

\begin{IEEEkeywords}
UAM fleet operation; UAM demand modeling; demand uncertainty; integer program
\end{IEEEkeywords}

\section{Introduction}
Urban Air Mobility (UAM) has been increasingly recognized for its potential to revolutionize urban transportation. It offers a novel alternative to the conventional modes of transportation that are troubled by issues such as congestion. As UAM adoption grows, effective fleet management strategies become crucial to ensure operational efficiency and high service quality, especially under demand uncertainty.

Since the mid-1970s , the airline industry has been using spill models to forecast average lost sales when demand surpasses flight capacity \cite{b1}. These models play a crucial role in informing critical decisions regarding aircraft configuration, fleet planning for growth, and fleet assignment to specific markets. Essentially, spill models have become a cornerstone of the contemporary yield management systems by framing demand for a group of flights within a probabilistic distribution. This strategic approach facilitates decision-making for airline operations, from single flight segments to complex networks serviced by various aircraft types \cite{b2} \cite{b3}.

Spill management, therefore, is pivotal in navigating the challenges posed by the mismatch between passenger demand and available capacity. In scenarios where the demand exceeds the aircraft's capacity, some passengers become "spilled" – they are either unable to fly or are rerouted through alternative flights offered by the airline. Spill estimates are used in short to medium-term airline planning decisions, such as fleet scheduling and assignment, and in longer-term fleet planning and strategic planning \cite{b4}. The correct estimation of spill costs is particularly important to fleet assignment.

As we pivot from traditional airline operations to the burgeoning field of UAM, the principles of spill management take on new dimensions and complexities. Urban environments, characterized by their dynamic and often unpredictable demand patterns, present a unique set of challenges for UAM operators. Unlike airlines, which can rely on extensive historical data to forecast demand, UAM initiatives must navigate a landscape where demand can fluctuate dramatically due to factors such as urban events, weather conditions, and evolving transportation trends.

\section{Related Work}
UAM is an exciting development in aviation technology as electric vertical take and landing (eVTOL) aircraft are used for fast city transport. However, the success of UAM depends on the ability to overcome many operational and infrastructure challenges \cite{b9} \cite{b8} . The key factor to the efficiency of UAM is the fleet management strategy, characterized by flight scheduling and the charging policy.

\subsection{Fleet Size Management}
Lots of research has been done to determine the best setup for UAM systems. \cite{b10} analyzed the performance of the UAM system under different demand and examines the impact of fleet and vertiport sizing has on ensuring the system's performance. \cite{b11} developed a model to minimize the costs of operating UAM systems by choosing the optimal eVTOL battery capacity and the power rate of the eVTOL chargers. Moving away from searching for the optimal solution, \cite{b12} used simulations to find the least number of aircraft needed, assuming that there is no limit to the number of aircraft and that the chargers have a power of 50 kW. For more specific strategies, another study \cite{b13} created charging plans in a port to reduce total waiting time of passengers.

\subsection{Fleet Assignment}

There has been much improvement in different areas of managing how the UAM systems handle unexpected situations.\cite{b11} used a bipartite matching dispatcher to assign aircraft to flights with the help of online simulation. \cite{b15} explored various methods to control aircraft approaches. \cite{b16} represented the network of vertiports as a graph and applied reinforcement learning to assign aircraft to flights, making sure that both charging needs and demand are met.  \cite{b17} looked into using a mix of different types of aircraft and developed a genetic algorithm to create nearly perfect schedules. There are also ongoing efforts focused on controlling the order in which the aircraft arrive \cite{b18}.

Although many studies have addressed the issues of charging limits and unpredictable demand in UAM systems, a unified approach to managing operations is lacking. Past research has not fully considered how charging power changes or how charging times should be coordinated with flight schedules. The analysis spill has also been largely ignored in the UAM realm. Our goal is to fill these gaps by proposing a more detailed method for planning UAM operations. We propose a two-stage method that aims to not only reduce the number of aircraft needed, but also to minimize disruptions during the assignment of flights. We apply a realistic, non-linear charging model to develop charging strategies that work jointly with flight scheduling and take into account the uneven and uncertain demand.

\section{Contributions}

In this work, we make two key contributions:
\begin{enumerate}
    \item \textbf{Stochastic UAM demand generation:} We construct a UAM demand model that generates stochastic passenger arrivals based on real-world data for a two-vertiport network considering both the day-to-day variation and the within-the-day variation in demand.
    \item \textbf{Optimization models for analyzing the trade-off between fleet size and spill under uncertain demand:} We propose two integer programs that model the dynamic nature of vertiport operations and the non-linear charging time of eVTOL aircraft. The two optimization models compute the zero-spill fleet size and the spill-minimizing operation policies, respectively, shedding light on UAM operations under uncertain demand.

\end{enumerate}

\section{Methodology}

\begin{figure}[t]
\centering
\includegraphics[width=0.5\textwidth]{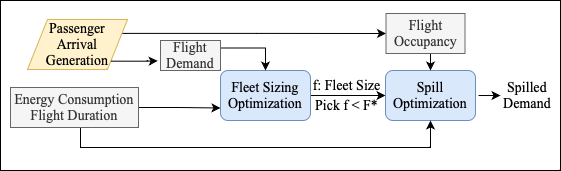}
\vspace{-0.8cm}
\caption{Analytical Framework}
\label{fig:framework}
\end{figure}

Our analytical framework consists of three modules, as shown in fig. \ref{fig:framework}. The first module is a stochastic process that generates UAM passenger arrivals at vertiports, capturing various sources of variation in demand. We derive the flight demand and flight occupancy from the passenger arrival distributions, which are the input parameters in the two optimization models. Fleet sizing optimization, the second model in the framework, computes the minimum number of aircraft required to fully satisfy the demand output from the demand generation module. Lastly, having computed the spill-minimizing fleet size, we choose a fleet size that is smaller than the spill-minimizing fleet size to compute the spill resulted from this less than optimal fleet size. 

\subsection{Stochastic Demand Modeling} \label{sec:model_demand}

In a two-vertiport system, we assume one vertiport is located in the Central Business District (CBD) of a city and the other vertiport is located at the major airport (APT) of the urban region. The demand for UAM service is driven by airline traffic at the airport. We assume airline arrivals result in demand for UAM from APT to CBD, and airline departures create demand from CBD to APT. We capture both the day-to-day and within-the-day variation using the UAM passenger arrival process generator outlined in algorithm \ref{algo1}.

For convenience, we introduce the following notations. Let $D$ be a set of airline schedules at APT of different days, $H=\{0,1,\dots,23\}$ denote the 24 hours of a day, and $V\in\{0,1\}$ be a set of flight directions where 0 indicates an airline flight arrival at APT and 1 indicates an airline flight departure at APT. Let $f_{d,h,v}$ be the set of airline flights on day $d \in D$, hour $h \in H$, and of direction $v \in V$. We use different definitions of $t$ for flights of different directions. For airline flights arriving at APT, we use the actual gate-in time as the definition of $h$. For airline flights departing from APT, we use the scheduled departure time as the definition of $h$. We define $\lambda_{i}$ as the expected number of passengers on airline flight $i$ that would use UAM to travel between APT and CBD. We use $c_{i}$ to denote the seat capacity of flight $i$. We further define $F_{d,h,v}(x)$ as the empirical cumulative distribution function (ECDF) of the seat capacity of flights on day $d$, in hour $h$, and of direction $v$, thus capturing the relative distributions departure or arrival of seats at APT. We define expected average daily directional demand ($ADD$) as the total expected number of passengers that travel between APT and CBD in each direction. Used as a parameter in the demand model, we can generate the passenger arrivals at different demand levels.

\begin{algorithm}
	\caption{Autoregressive Passenger Arrival Process}
	\begin{algorithmic}[1]
        \State $ADD$ = Expected average daily directional demand
        \State $\beta_{v} = \sum_{d}\sum_{h}\sum_{i \in f_{d,h,v}} c_{i} $
        \For{$d\in D$}
            \For{$h \in H$}
                \For{$v \in V$}
                    \State $\lambda_{i}$ =  $\frac{c_{i}}{\beta_{v}}\cdot ADD \cdot |$D$|$ $\forall i\in f_{d,h,v}$
                    \State $\Lambda_{d,h,v}^{0} = \sum_{i \in f_{d,h,v}} \lambda_{i}$
                    \State $\Lambda_{d,h,v} = \Lambda_{d,h,v}^{0} + [(x_{d, h-1. v}-\Lambda_{d, h-1, v}^{0})\cdot\frac{\Lambda_{d,h,v}^{0}}{\Lambda_{d, h-1, v}^{0}}]\alpha$ 
                    \State $x_{d,h,v} = Pois(\Lambda_{d,h,v})$
                    \For{$p$ in $x_{d,h,v}$}
                        \State Generate $u$ from $Unif(0,1)$
                        \State Assign passenger $p$ to flight $i\in f_{d,h,v}$ using $F^{-1}_{d,h,v}(u)$
                        \State Generate the lag/lead time for each passenger from the skewed normal distribution
                        \State Add lag/lead time to the time of arrival/departure of the flight
                    \EndFor
                \EndFor
            \EndFor
        \EndFor
        
	\end{algorithmic}
    \label{algo1}
\end{algorithm}

\subsubsection{Day-to-Day Variation}

Because airline schedules exhibit temporal variation across different days of the week, seasons, and special occasions such as holidays, we capture this day-to-day variation by first calculating the total yearly seat capacity of flights at APT, $\beta_{v}$, and then allocating the yearly UAM demand proportionally to each individual flight, as shown on line 6 in algorithm \ref{algo1}. Therefore, on days with higher airline passenger volume at APT, which can be attributed to both the increase in number of flights and the use of larger aircraft, we expect to see more UAM passengers as well. 

\subsubsection{Within-the-Day Variation}

We capture within-the-day variation in passenger arrivals through autoregression. We assume that on days when we observe higher or lower demand than the average early on, we expect to have higher or lower demand throughout the day. We model autocorrelation as a function of the expected number of passengers in the current and the previous hour, and the realized number of passengers in the previous hour. As shown in line 8 of algorithm \ref{algo1}, the actual rate of arrivals, $\Lambda_{d,h,v}$, is updated based on both belief and observations for each hour. We use the autoregressive coefficient $\alpha \in [0,1]$ as a parameter that controls the degree to which consecutive demand rates are correlated.  When $\alpha=0$, we only consider the day-to-day variation that exists inherently in the airline schedules and assume no autocorrelation. 

After obtaining an hourly rate, $\Lambda_{d,h,v}$, we generate the number of passengers with a Poisson process and allocate the observed number of passengers to each flight $i\in f_{d,h,v}$ based on the inverse CDF of the seat capacity in each hour. This allows us to obtain a UAM passenger count for each airline flight. To provide realistic passenger arrival time at vertiports, we sample the lead or lag time for UAM passengers from skewed normal distributions. For UAM passengers departing from the urban center for the airport, we define their arrival time at the CBD vertiport as the departure time of the airline flight from the airport minus the lead time. For UAM passengers arriving at the airport and going to the urban center, we define their arrival time at the APT vertiport as the time of the airline flight arriving at the gate plus the lag time. 

The 2013 Airport Cooperative Research Program (ACRP) report is used to estimate the lead and lag time distributions \cite{b19}. The lead time follows a skewed normal distribution with $\mu=93$, $\sigma=40$, and the skewness parameter $\alpha = 3$, which results in a median lead time of 120 minutes \cite{b19}. The ACRP report does not provide a distribution for lag time. By taking the provided mean lag time, 21 minutes, and assuming a transfer time of 10 minutes from the curb of the airport to the vertiport, we arrive at a skewed normal distribution with $\mu=31$, $\sigma=2.12$, and $\alpha = 3$ \cite{b19}.

\begin{figure}[t]
\centering
\includegraphics[width=0.5\textwidth]{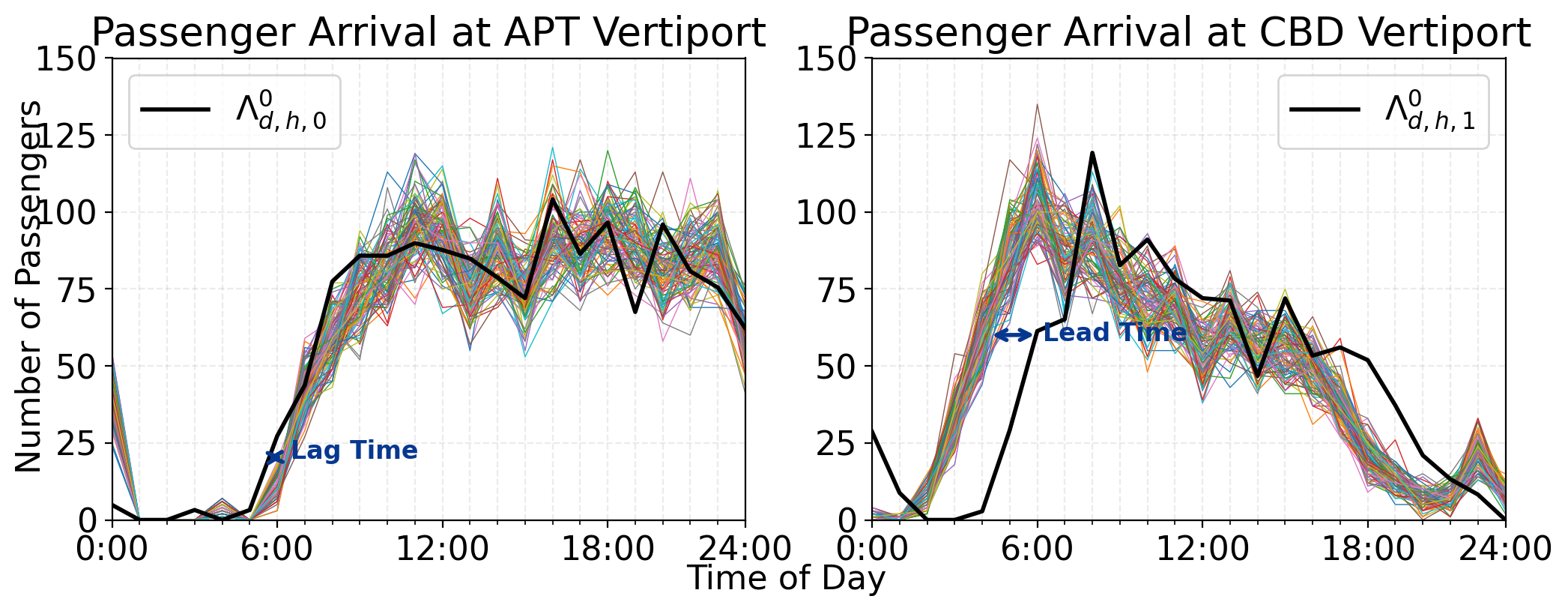}
\vspace{-0.8cm}
\caption{An example of 100 realizations of autoregressive passenger arrival time generation using the flight schedule of a major U.S. airport on 1/1/2019. The autoregressive coefficient is set to $\alpha=0.7$. The realized passenger arrival process vary around the expected hourly rate $\Lambda^{0}_{d,h,v}$ and the gap between the realized trajectories and the expected rate is due to the addition and subtraction of lead and lag time.}
\label{fig:demand}
\end{figure}

\subsection{Optimization Models}
To understand the impact that variation in demand has on fleet operation and fleet sizing, we adopt a two-step process by which we first compute the minimum fleet size required to fully satisfy a certain demand profile and then compute the spill with a smaller than minimum fleet size. We define spill as the demand denied of service by the UAM operator, in number of passengers. By solving these two integer programs, we are able to analyze the trade-off between fleet size and spill. The notation for the optimization models can be found in table \ref{tab:notations}. Notations that are written as a function of time are decision variables and the rest are model parameters computed from model inputs.

\subsubsection{Fleet Sizing (Zero-Spill) Optimization}\label{sec:op1}

In the first integer program, we find the minimum number of aircraft required to satisfy a given flight demand profile, generated using the stochastic passenger arrival process outlined in section \ref{sec:model_demand}. We consider a finite set of discretized time intervals at which control actions can be taken. We also consider a finite set of State of Charge (SoC) levels to model the battery state of the eVTOLs. An aircraft can be in one of three states: idling, charging, or in flight. As the decision variables dictate the transition of aircraft among these states, we are able to model the dynamics of the UAM system. We define the objective function as the following:
\vspace{-1mm}
\begin{align}
\min &\sum_{i}\sum_{k} n_{i}^{k}(t=0) + \sum_{i} \sum_{x} \sum_{y} C_i^{xy}(t = 0) \nonumber \\
     &+ \alpha \cdot  \sum_{t} \sum_{i} \sum_{j} \sum_{k} u_{ij}^{k}(t) 
\end{align}

The first two summation terms in the objective function represent the total number of eVTOL aircraft, based on time step $t=0$, i.e., at the beginning of an operating day. Because no aircraft is flying at $t=0$, the sum of aircraft idling and charging at $t=0$ constitutes the fleet size. The third term in the objective function represents the total number of flights with a small weight $\alpha$, which has the purpose of constraining the solution space so that between two solutions that yield the same fleet size, the solution that minimizes the number of flights is preferred. We set $\alpha = 0.00001$ because $\alpha$ has to be small enough such that $\alpha \cdot \textit{total number of flight}$ does not exceed 1.

\begin{table}
    \centering
    \caption{Optimization Model Notations}
    \renewcommand{\arraystretch}{1.5} 
    \begin{tabular}{c|p{6.6cm}}
     \textbf{Notation} & \parbox{5cm}{\centering \textbf{Explanation}}   \\ \hline 
     $C_{i}^{xy}(t)$ &  Number of aircraft at vertiport \(i\) that begin to charge at time step \(t\) with an initial SoC of \(x\) and a target SoC of \(y\)  \\ \cline{1-2}
     \(u_{ij}^{k}(t)\) &  Number of aircraft departing for vertiport \(j\) from vertiport \(i\) at time step \(t\) with SoC \(k\)  \\ \hline
     
     \(n_{i}^{k}(t)\) &  Number of idle aircraft at vertiport \(i\) of SoC \(k\) at time step \(t\)  \\ \cline{1-2}

    $s_{ij}(t)$ & Number of passengers spilled travelling from vertiport $i$ to $j$ at time step $t$ \\ \cline{1-2}

    $p_{ij}^{t}$ & Demand in number of passengers travelling from vertiport $i$ to $j$ at time step $t$ \\ \cline{1-2}

     \(f_{ij}^{t}\)  &  Number of flights departing from vertiport $i$ to vertiport $j$ at time step $t$ required to satisfy travel demand  \\ \cline{1-2}
     \(\gamma_{k}\) & Charging time needed to transition from \(SoC_{k-1}\) to \(SoC_{k}\)  \\ \cline{1-2}
     
     \(\tau_{ij}^{t}\) & Flight time from vertiport \(i\) to \(j\) at time step $t$ in time steps  \\ \cline{1-2}
     \(\kappa_{ij}^{t}\) & Reduction in SoC for a flight from vertiport \(i\) to \(j\) departing at time step $t$  \\ \cline{1-2}

     $\mathcal{V}$ & Set of vertiports  \\ \cline{1-2}

     $\mathcal{A}_{ij}^{t}$ & $\{t^\prime \in \{1, \dots, T\} |  t^\prime+\tau_{ij}(t^\prime) = t\}$ \\ \cline{1-2}
     $T$ & Planning horizon $T = \text{number of time steps} + \max\limits_{ijt} \tau_{ij}(t) + 1$ \\ \cline{1-2}
     \(K\) & Number of SoCs after discretization  \\ \cline{1-2}
     $F$ & Fleet Size \\ \cline{1-2}

     $O$ & Seat capacity of the eVTOL aircraft \\

    \end{tabular}
    \label{tab:notations}
\end{table}
\vspace{-1mm}

Minimization of the objective function is subject to the following constraints:
\begin{align}
    &n_{i}^{k}(t) =  n_{i}^{k}(t-1) + \sum_{j\in\mathcal{V}-\{i\}} \sum_{t^\prime \in \mathcal{A}_{ji}^{t}} u_{ji}^{k+\kappa_{ji}^{t^\prime}}(t^\prime) - \sum_{j\in\mathcal{V}-\{i\}} u_{ij}^{k} (t) \notag\\
    & + \sum_{x=0}^{k-1} C_{i}^{xk} (t-\sum_{i=x+1}^{k} \gamma_{i}) - \sum_{y=k+1}^{K} C_{i}^{ky}(t), \quad \forall i,t 
     \label{cons:dynamic_equation}
\end{align}
\vspace{-2mm}
\begin{align}
    &\sum_{k \in\{1, \cdots, K\}} u_{ij}^k(t) \geq f_{ij}^{t}, \quad \forall i, j, t 
    \label{cons:demand}
\end{align}
\vspace{-4mm}
\begin{align}
    u_{ij}^0(t) = 0, \quad \forall i,j,t 
   \label{cons:energy}
\end{align}
\vspace{-5mm}
\begin{align}
    &n_{i}^{k}(0)=n_{i}^{k}(T), \quad u_{ij}^{k}(0)=u_{ij}^{k}(T), \quad C_{i}^{xy}(0)=C_{i}^{xy}(T) \quad  \nonumber \\  & \forall i,j,k,x,y
    \label{cons:sta}
\end{align}

\vspace{0.3em}

The dynamic equation that governs the evolution of the system is given by constraint (\ref{cons:dynamic_equation}), which models the state of the aircraft in relation to the charging policy and the flight schedule. We model the number of aircraft that enter a state of idle at each vertiport at each time step of a certain SoC as a linear combination of the following terms: (A) the number of idling aircraft that are carried over from the previous time step, (B) the number of aircraft that will arrive at vertiport $i$ at time step $t$ of SoC $k$, (C) the number of aircraft that depart from vertiport $i$ of SoC $k$, (D) the number of aircraft that complete charging to SoC $k$ at time step $t$ at vertiport $i$, and (E) the number of aircraft that are committed to charging at vertiport $i$ starting from SoC $k$.  Constraint (\ref{cons:demand}) ensures that flight demand for each vertiport pair $(i,j)$, measured as the required number of flights $f_{ij}^{t}$, is satisfied at each time step. Constraint (\ref{cons:energy}) ensures that aircraft cannot fly when the SoC level equals 0, which is the reserve SoC. Constraint (\ref{cons:sta}) ensures that the number of aircraft in different states and SoCs are the same at the beginning and the end of the time horizon. These constraints are essential to ensuring the feasibility of the flight schedule and charging policy.

\subsubsection{Spill Optimization}

Although an optimal (zero-spill) number of aircraft is obtained from the program in section \ref{sec:op1}, serving the same demand with fewer than optimal number of aircraft does not produce linear increase in spill, i.e, the flight schedule and charging policy can be updated to serve similar number of passengers with a smaller fleet size. After obtaining the optimal number of aircraft, $F^{*}$, from the first optimization model, we choose a fleet size $f \in \{F | F < F^{*}\}$ to compute the spill given such fleet size $f$. We formulate the spill optimization problem as:

\begin{align}
\min \sum_{t}\sum_{i} \sum_{j} s_{ij}(t) + \alpha \cdot  \sum_{t} \sum_{i} \sum_{j} \sum_{k} u_{ij}^{k}(t) \nonumber
\end{align}

where the first term is the sum of all spill over a day in number of passengers and the second term is the small penalty on the number of flights as discussed in the fleet sizing program. 

The spill optimization model retains constraint (\ref{cons:dynamic_equation}), (\ref{cons:energy}), and (\ref{cons:sta}) of the first program and we add three additional constraints. Constraint (\ref{cons:posi}) ensures that spill is a non-negative number. When Constraint (\ref{cons:posi}) is not binding, Constraint (\ref{cons:maxspill}) computes the correct spillage by taking the difference between the total directional demand, in number of passengers, at time step $t$ and the total seat capacity offered on the flights scheduled. Constraint (\ref{cons:fs}) sets the fleet size.

\begin{align}
    s_{ij}(t) \geq 0, \quad \forall i,j,t 
\label{cons:posi}
\end{align}
\begin{align}
    s_{ij}(t) \geq p_{ij}^{t} - O \cdot \sum_{k} u_{ij}^{k}(t) \quad \forall i,j,t
\label{cons:maxspill}
\end{align}

\begin{align}
    \sum_{i}\sum_{k} n_{i}^{k}(t=0) + \sum_i \sum_{x} \sum_{y} C_i^{xy}(t = 0) = F
\label{cons:fs}
\end{align}

\section{Numerical Experiment}

We demonstrate the implementation of our analytical framework with a two-vertiport network. We discretize a day into 288 time steps on which charging, idling, and scheduling decisions are made. We also discretize the aircraft battery capacity into 32 levels of SoCs above the 20\% reserve requirement, each in 2.5\% increment. We assume a non-time varying directional energy consumption of 4 levels of SoC (10\%) and a non-time varying flight time of 2 time steps (10 minutes) between the two vertiports. 

\begin{figure}[t]
\centering
\includegraphics[width=0.4\textwidth]{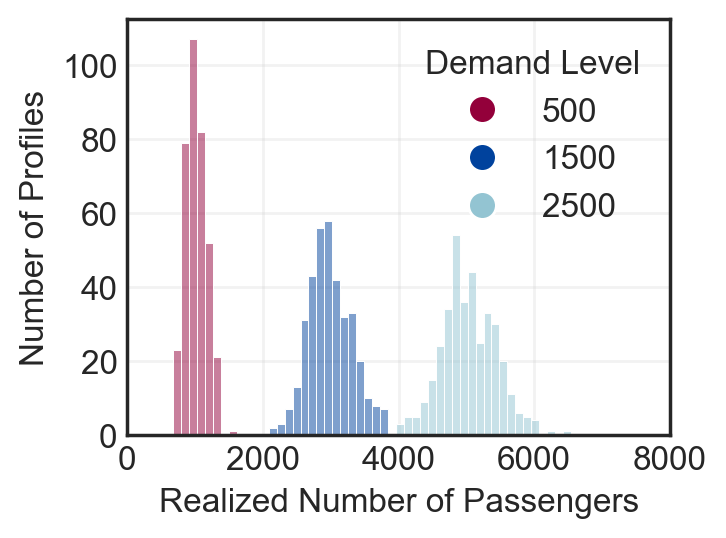}
\vspace{-0.5cm}
\caption{Distribution of daily realized number of passengers across different demand levels ($ADD$)}
\label{fig:realized_passengers}
\end{figure}

We obtain the daily airline flight schedules of a major U.S. airport in the year 2019 from the Aviation System Performance Metrics (ASPM) database provided by the Federal Aviation Administration (FAA) \cite{aspm}. We obtain the aircraft seat capacity through the Form 41 Air Carrier Statistics database from the Bureau of Transportation Statistics \cite{bts}. We generate passenger arrival process at three demand levels: 500, 1500, and 2500, which are the expected average daily directional demand (ADD) defined in section \ref{sec:model_demand} with the autoregressive coefficient $\alpha$ set to 0.7. We also generate an additional set of demand profiles at $ADD=1500$ and an autoregressive coefficient $\alpha = 0$ to study the effect within-the-day variation produces. For each demand level, we create one realization per day in the airline flight schedules, totalling 365 passenger arrival profiles. Fig. \ref{fig:realized_passengers} shows the distribution of the realized number of passengers across the profiles. We observe larger variation in the realized number of passengers as the demand level rises, due to the nature of the Poisson process used in which the variance of the random variable is equal to its expected value.

\begin{figure}[t]
\centering
\includegraphics[width=0.5\textwidth]{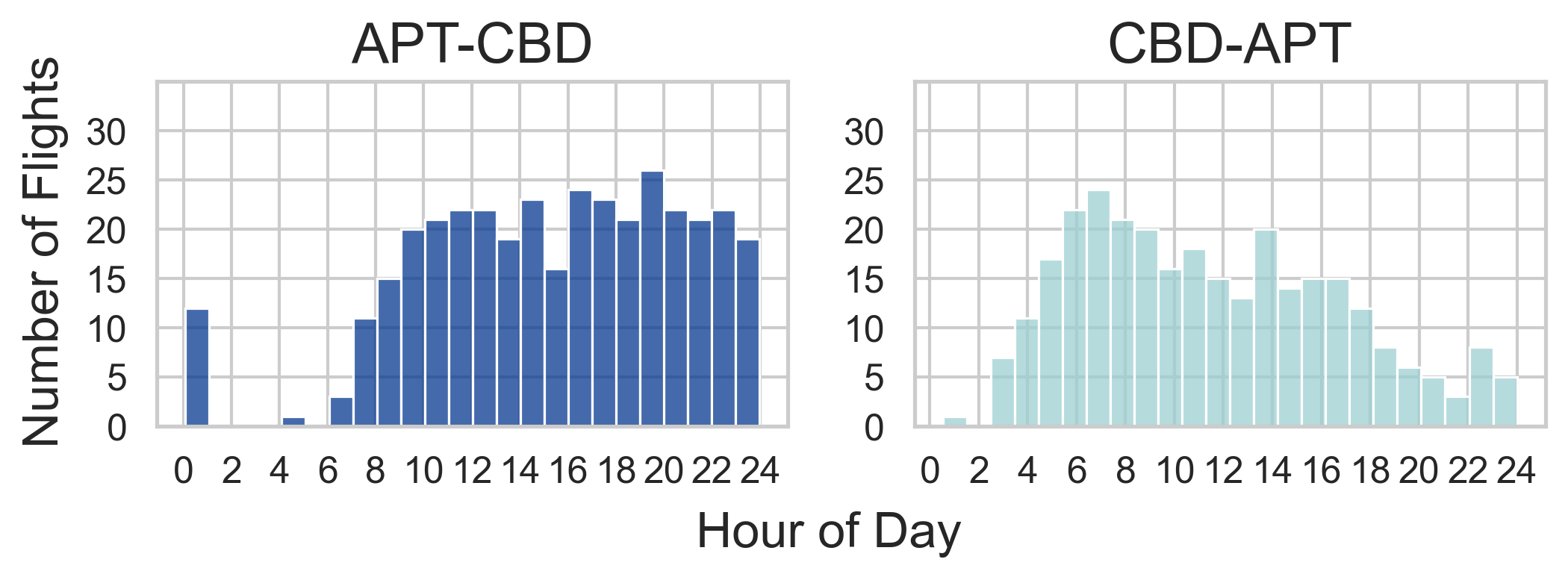}
\vspace{-0.8cm}
\caption{Aggregated hourly flight demand on one realization of passenger arrivals after the conversion from passenger arrival to flight demand using the 5-minute dispatch policy.}
\label{fig:schedule_demand}
\end{figure}

We adopt a deterministic flight dispatching policy to convert the passenger arrival process outlined in section \ref{sec:model_demand} to flight demand $f_{ij}^{t}$ used in the fleet sizing program. In this conversion, we also compute the corresponding occupancy of these flights, which are then used to calculate $p_{ij}^{t}$, used in the spill optimization. The deterministic flight dispatch control creates a flight demand if $O$ (occupancy of the eVTOL aircraft) passengers have arrived at a vertiport or if the first passenger has waited in queue for more than 5 minutes. With this dispatch rule, we compute the flight demand as shown in Fig \ref{fig:schedule_demand} for each passenger arrival profile.

\section{Results and Discussion}

\subsection{Fleet Sizing}

\begin{figure}[h]
\centering
\includegraphics[width=0.4\textwidth]{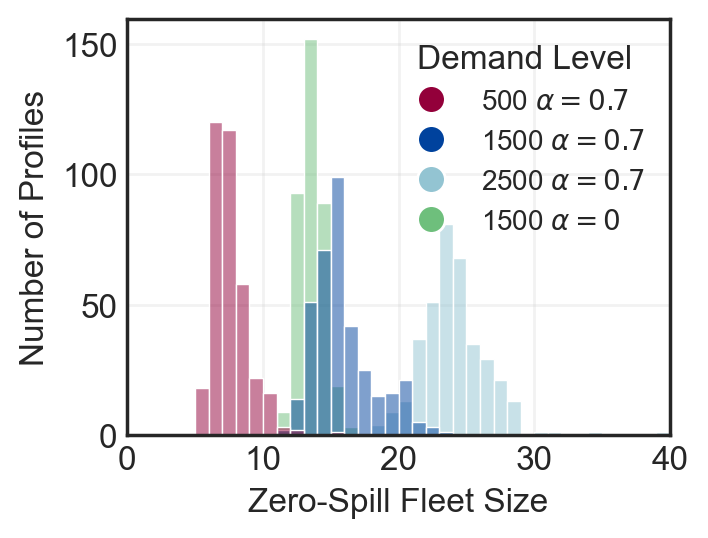}
\vspace{-0.5cm}
\caption{Distribution of zero-spill fleet size computed by the fleet sizing optimization model for different sets of demand model parameters.}
\label{fig:fs}
\end{figure}

As fig. \ref{fig:fs} reveals, not only do higher demand levels lead to a larger fleet size, they also result in larger variance in fleet size. However, if we do not consider the within-the-day variation captured by the autoregressive process, the range of zero-spill fleet size decreases from 10 to 5, as indicated by the difference between the two sets of profiles generated at an $ADD$ of 1500 but of different $\alpha$, the autoregressive coefficient. When we consider within-the-day variation in demand, more aircraft are needed to satisfy the demand. On the other hand, by the design of the passenger arrival process generator, the relatively small variance in fleet size for $\alpha=0$ indicates that the day-to-day variation in airline schedules plays a less significant role in fleet sizing. 

\subsection{Demand Spillage}

\begin{figure}[h]
\centering
\includegraphics[width=0.45\textwidth]{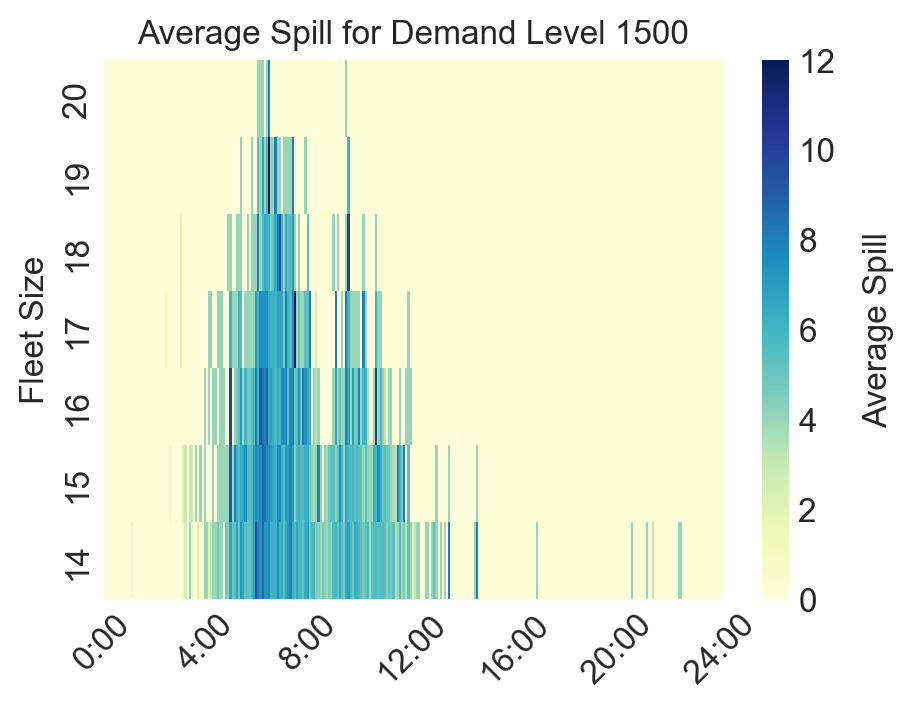}
\vspace{-0.5cm}
\caption{Temporal distribution of spill across different fleet size at a directional demand level of 1500 with autoregressive coefficient $\alpha$ at 0.7.}
\label{fig:spill_1500}
\end{figure}

Fig. \ref{fig:spill_1500} shows the temporal distribution of the average spill over all demand profiles generated at $ADD = 1500, \alpha=0.7$ across different sub-optimal fleet sizes. The result shows that at higher fleet size, spill only occurs in the early morning, when the difference in flight demand between the two directions is the greatest, as shown in fig. \ref{fig:demand}. As the fleet size decreases, the spill optimization model produces solutions that spill passengers in other periods during the day, and the time span of spill increases. This result suggests that the driving factor behind spill is not the high flight demand but rather the high imbalance in demand among different flight directions. 

To analyze the trade-off between fleet size and service quality (i.e. spill) for the two-vertiport system, we compute the average daily spill in number of passengers at different fleet sizes. We define spill over a day as $\sum_{i}\sum_{j}\sum_{t} s_{ij}(t)$ and the average daily spill as the mean of all spill over a day in the one-year period. To further validate the solution and provide context, we use two sets of heuristics to derive the upper and lower bound for the average daily spill.  

\subsubsection{Upper Bound} We make the following assumptions in generating the upper bound:
\begin{itemize}
    \item An aircraft is committed to charge when its SoC falls to the reserve level or if it is idling.
    \item Within the same time step, aircraft first serve flights that have a larger number of passengers.
    \item There are no repositioning flights except for one scenario. When one vertiport has more than 5 aircraft of an SoC of 50\% or more idling, we send a sufficient number of aircraft to the opposite vertiport such that we have at max 5 aircraft at 50\% or more idling. 
\end{itemize}

\subsubsection{Lower Bound} We make the following assumptions in generating the lower bound:
\begin{itemize}
    \item An aircraft is committed to charge when its SoC falls to the reserve level or if it is idling.
    \item Within the same time step, aircraft first server flights that have a larger number of passengers.
    \item We combine the two vertiports into one demand and server. As soon as an aircraft finishes its previous flight, it is ready to serve passengers again regardless of the location of the aircraft given that it has sufficient SoC. With such assumption, we eliminate the impact of repositioning flights.
\end{itemize}

\begin{figure}[t]
\centering
\includegraphics[width=0.5\textwidth]{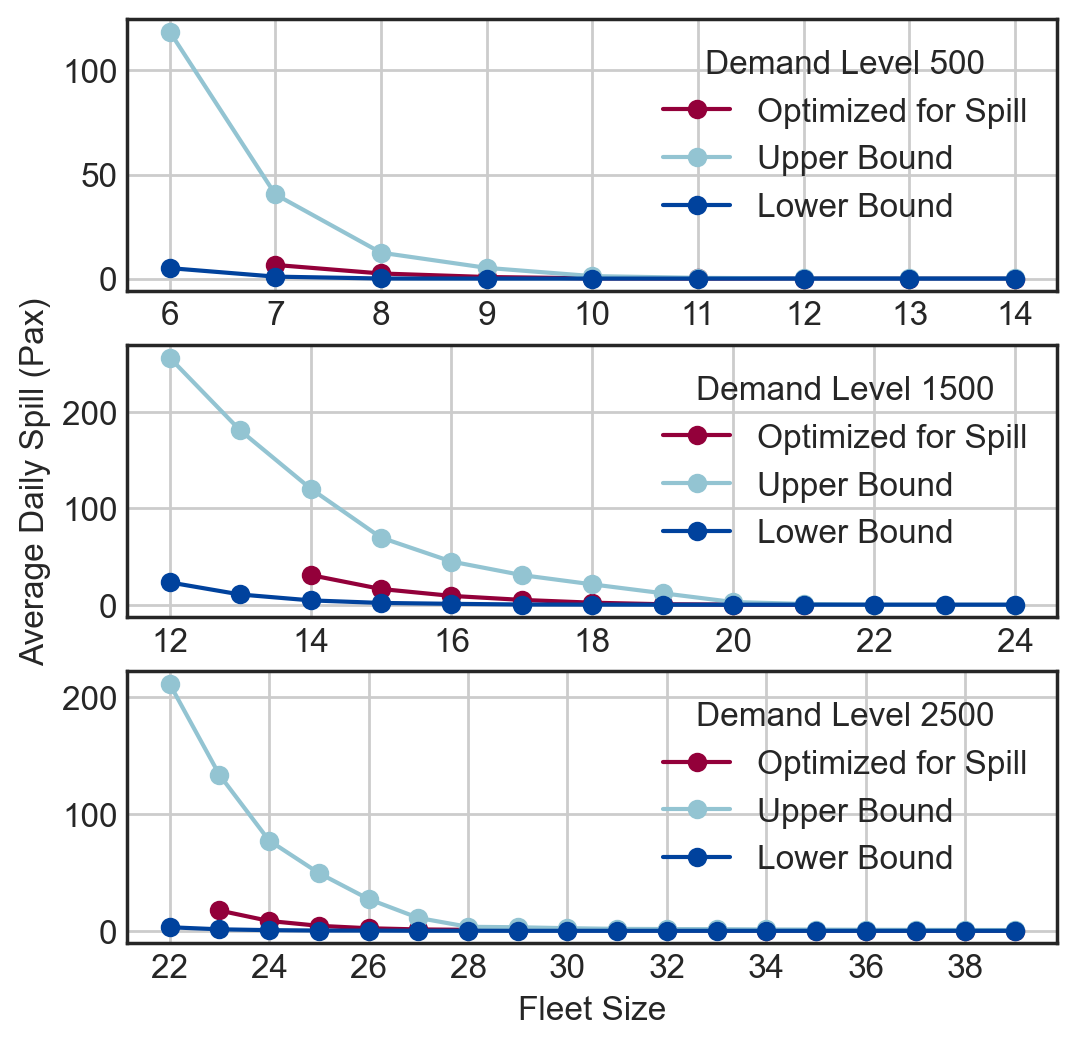}
\vspace{-0.8cm}
\caption{Average daily spill at different fleet sizes.}
\label{fig:trade}
\end{figure}

With these heuristics, we simulate the operation given the generated demand profiles. At each time step, we sum the occupancy of the flights that are not served. Fig. \ref{fig:trade} shows the trade-off between the average daily spill and the fleet size. We observe a convex relationship between all three spill metrics and fleet size. The results reveal that spill is relatively inelastic to fleet size. For example, we know the median of the zero-spill in fleet size for the $ADD=1500$ scenario is 15, shown in fig. \ref{fig:fs}. However, at this median fleet size, we observe an average daily spill less than 40 passengers, which counts for less than 2\% of the total daily demand of 3000. As we continue to reduce the fleet size parameter, $F$, in the spill optimization model, the solver experience long running time in reducing the optimality gap. Therefore, we do not compute the average daily spill at small fleet sizes. Still, it shows the importance of optimizing the charging policy and the flight schedules for different demand patterns while retaining the fleet size. In a real world setting, dynamically adjusting to demand can potentially reduce the capital cost of acquiring aircraft substantially. 

\begin{figure}[t]
\centering
\includegraphics[width=0.5\textwidth]{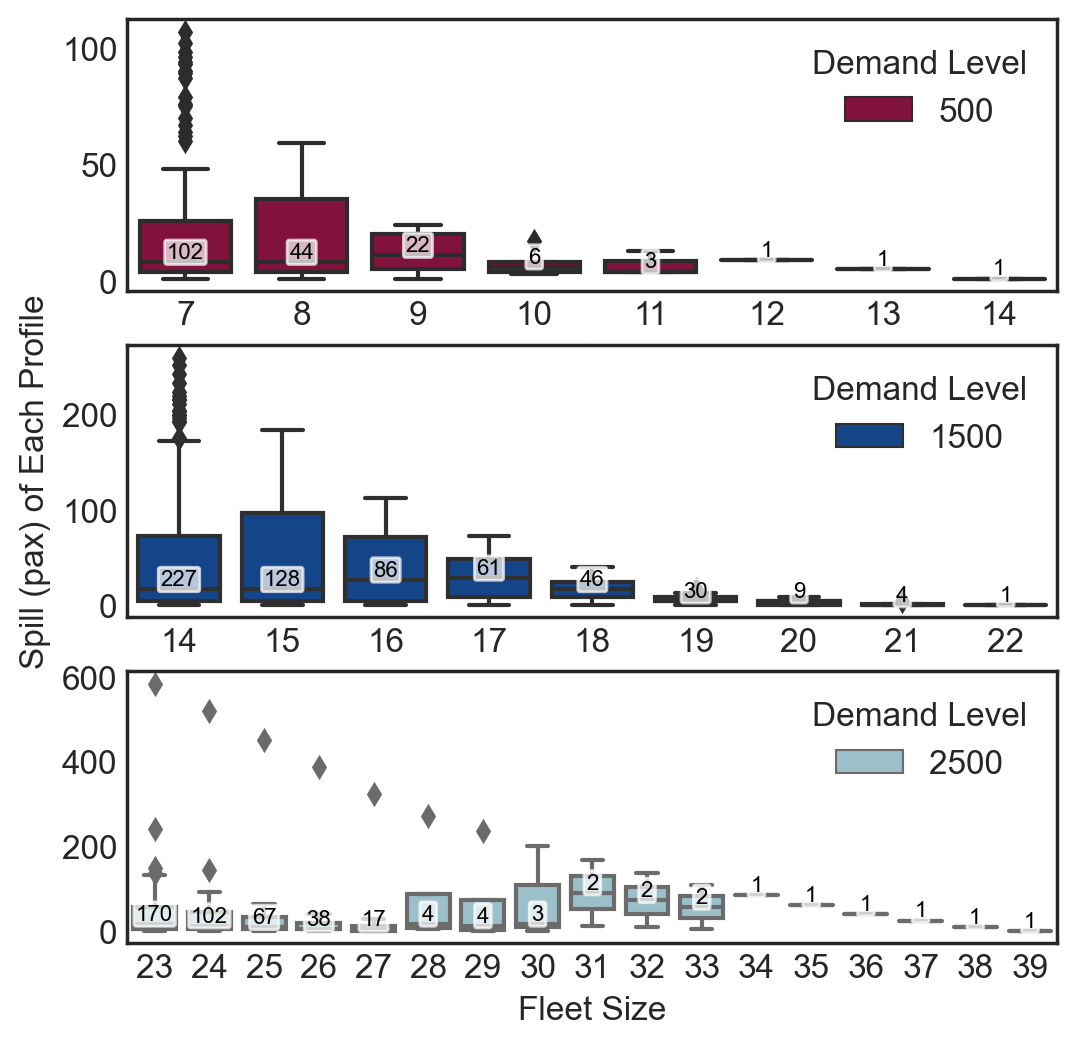}
\vspace{-0.8cm}
\caption{Variance in spill across different fleet sizes. The number on each box indicates the sample size (i.e. the number of profiles that have a zero-spill fleet size smaller than $i+1$ where $i$ is the corresponding fleet size of the box.}
\label{fig:spill_var}
\end{figure}

Although choosing the median fleet size across all the profiles at each demand level does not lead to high spill levels on average, the variance in spill, however, is much greater. Fig. \ref{fig:spill_var} shows the variance in spill at each fleet size for different demand levels. As the fleet size decreases, we observe a drastic increase in the number of high-spill days, signaling potential service disruptions. Additionally, as the demand level $ADD$ increases, the system appears more robust against demand variation. At an $ADD$ of 500, choosing a fleet size, 7, that is slightly higher than the median produces many days of spilling around 100 passengers, which counts for 10\% of the total expected daily demand. In contrast, when $ADD$ is 2500 and that we choose the median fleet size of 23, we only have 2 days that result in spill over 200 passengers, which is less than 5\% of the expected total demand. This suggests maintaining a larger fleet and serving more passengers on average gives operator more flexibility in designing charging policies and flight schedules to minimize spill.

\section{Conclusion}

Our paper explores the relationship between fleet size and spill, which is the number of passengers turned away from service, under demand uncertainty. We build a stochastic demand model that generates UAM passenger arrival process based on real-world data, a fleet sizing optimization model that computes the zero-spill fleet size, and a spill optimization model that designs flight schedules and charging policies that maximize the number of passengers the UAM system can serve. Our research finds that spill is relatively inelastic to fleet size, especially at high demand levels. We also conclude that the driving factor behind spill is the demand imbalance between different origin-destination pairs in the UAM system. 

In the future, we aim to adopt heuristic and relaxation techniques for solving the two integer programs to reduce running time. This would allow us to analyze beyond a two-vertiport network and spill at a smaller fleet size. We also seek to leverage reinforcement learning (RL) and VertiSim, a vertiport operation simulator, to better model the details of fleet operations, such as air holding time, taxing time, and variation in charging time. A comparison in the charging and scheduling policies output between the optimization models and the RL agent would help us better understand the intricacy of UAM vertiport and fleet operations.

\vspace{12pt}

\end{document}